\documentclass{webofc}

\usepackage[varg]{txfonts}   
\usepackage{hyperref}
\usepackage{url}
\hypersetup{colorlinks=true,citecolor=blue,urlcolor=blue,linkcolor=blue}
\usepackage{rotating}
\usepackage{xcolor}
\usepackage{amsmath}

\usepackage[utf8]{inputenc}

\usepackage[section]{placeins}

\usepackage{comment}
\usepackage{subfigure}
\usepackage{subcaption}

\usepackage{natbib}
    
\usepackage{lineno}

\begin{document}%
%
%

%
\title{Probing jet hadrochemistry in \mbox{Pb--Pb} collisions with ALICE}
%
%

\author{\firstname{Sierra} \lastname{Cantway}\inst{1}\thanks{\email{sierra.lisa.weyhmiller@cern.ch}} for the ALICE Collaboration
}

\institute{Yale University Wright Laboratory 272 Whitney Ave., New Haven, USA}

\abstract{Jet substructure measurements in heavy-ion collisions provide constraints on jet quenching and the medium response in the QGP. Though there has been remarkable progress in inclusive-charged-hadron jet substructure measurements, understanding the identified particle composition of jets and their modification in heavy-ion collisions has proven challenging. Jet quenching models predict that the jet hadrochemical composition may be modified in heavy-ion collisions due to jet-medium interactions, as well as the medium response. Measurements of identified particles in jets can help discriminate between various parton-QGP interactions. We present the first measurements of $\pi$, K, and p ratios within charged-particle-jets, as well as in the underlying event, as a function of particle transverse momentum in pp and \mbox{Pb--Pb} collisions at $\sqrt{s_{\rm{NN}}}$ = 5.02 TeV. These measurements leverage the excellent PID capabilities of ALICE over a wide transverse momentum range. These results aim to understand soft particle production mechanisms and distinguish modified jet fragmentation from bulk effects.}
\maketitle
\section{Introduction}
\label{intro}
Heavy-ion collisions create a high-energy-density state of matter called the Quark Gluon Plasma (QGP), where partons are no longer confined. Energetic partons traverse the QGP medium as it evolves and fragment into collimated showers of particles called jets. Interactions with the medium are expected to modify this parton shower through elastic collisions with medium partons~\cite{Bjorken1982} and inelastic medium-induced gluon radiation~\cite{GluonRad}. In addition, jets are expected to modify the medium itself by inducing a correlated wake response, where there is an enhancement of particles in the jet-going direction at large angles and a depletion of particles on the opposite side~\cite{WakeKrishna}. Measurements of the internal structure of these jets (jet substructure) can constrain how energetic partons interact with the QGP. There is a large body of jet substructure work aimed at achieving this goal, but much of it has focused on inclusive-charged-hadron jet substructure measurements \cite{JetReview2018}. As such, knowledge of the identified particle composition of the jet and its modification in the QGP remains incomplete.

Progress has been made towards this, however, as there have been some studies on light-flavor and heavy-flavor identified particles in jets in both theory and experiment. These proceedings focus on the light-flavor sector. For example, AMPT simulations predict that the particle composition in the jet can be modified in heavy-ion collisions due to parton coalescence in the jet wake \cite{Luo:2023}. Conversely, theoretical calculations predict a different modified particle composition due to enhanced parton splittings in the medium \cite{Sapeta:2007ad}. Measurements of identified particles in jets are thus sensitive to hadronization mechanisms, potential modified jet fragmentation, and potential wake contributions. There have been some experimental efforts towards this already. ALICE has previously found that baryon and strangeness production in jets is much lower than in inclusive production or production in the underlying event (UE) in pp and p--Pb collisions via measurements of $\Lambda$/$\rm{K^{0}_{S}}$ and $\Xi$/$\Lambda$ ratios \cite{CascadespPbpp}. Preliminary measurements from STAR have indicated that jets with a high-$\textit{p}_{\mathrm{T}}$ constituent requirement (track bias) do not exhibit baryon/meson modification in Au--Au collisions at \ensuremath{\sqrt{s_{\mathrm{NN}}}} = 200 GeV via p/$\pi$ ratios~\cite{GabeHP23Proceed}. However, track biases are known to bias measurements towards jets that are less quenched and towards a fragmentation pattern where one particle takes most of the jet $\textit{p}_{\mathrm{T}}$~\cite{JetReview2018}, both of which are expected to lead to less modified baryon/meson ratios. In these proceedings, ALICE presents measured K/$\pi$ and p/$\pi$ ratios in jets without a requirement on leading track momentum in both pp and \mbox{Pb--Pb} collisions to reduce these biases.

\section{Experimental Details}
\label{experiment}

This analysis is performed on 0$-$10\% centrality \mbox{Pb--Pb} data at $\sqrt{s_{\rm{NN}}} = 5.02$ TeV, taken with the ALICE detector \cite{ALICEDetector}. Only charged tracks with 0.15 $< $ $\textit{p}_{\mathrm{T}}$ $<$ 100 GeV/$c$ and -0.9 $< \eta <$ 0.9 are considered. There is no explicit $y$ requirement. Charged-particle jets are reconstructed using the anti-$k_{\rm{T}}$ algorithm and the $\textit{p}_{\mathrm{T}}$-scheme for recombination with a resolution parameter, $R$, of 0.4 \cite{Cacciari:2008gp}. Jets must be within the fiducial acceptance of the Time Projection Chamber (TPC) ($-0.5 < \eta_{\rm{jet}} < 0.5$). Jets are required to have a minimum area of $0.6\pi R^{2}$ to suppress background-only jets. The jet $\textit{p}_{\mathrm{T}}$ is corrected for the average event-wise background with the traditional pedestal subtraction method \cite{RhoSub}. Jets are required to have a jet $\textit{p}_{\mathrm{T}}$ after pedestal subtraction ($p^{\rm{raw~sub}}_{\rm{T,~ch~jet}}$) of 60\textendash140 GeV/$c$ to minimize the contribution of purely combinatorial (fake) jets. We note that $p^{\rm{raw~sub}}_{\rm{T,~ch~jet}}$ is not equal to the true charged-particle jet $\textit{p}_{\mathrm{T}}$ ($p_{\rm{T,~ch~jet}}$) due to the smearing of the jet $\textit{p}_{\mathrm{T}}$ from detector and residual background fluctuations. However, only a weak jet $\textit{p}_{\mathrm{T}}$-dependence is expected in the K/$\pi$ and p/$\pi$ ratios. As such, the final results are reported at the level of $p^{\rm{raw~sub}}_{\rm{T,~ch~jet}}$. Any potential impact is accounted for in the results' systematics by varying the minimum $p^{\rm{raw~sub}}_{\rm{T,~ch~jet}}$ $\pm$5 GeV/$c$ to vary the truth jet $\textit{p}_{\mathrm{T}}$ populations. 

Pedestal subtraction does not remove the background constituents of the jets themselves. The bulk underlying event contribution dominates the jet signal at low $\textit{p}_{\mathrm{T}}$, even inside the jet cone. Thus, data-driven, species-dependent underlying event subtraction is needed to probe the particle production mechanisms associated with the hard scattering. In this measurement, this is achieved via particle identification (PID) measurements performed on different particle sources, which are then used to perform UE subtraction. The jet cone particle source (JC) is defined as all the particles inside the selected anti-$k_{\rm{T}}$ reconstructed jet cones, i.e. both the jet signal particles and the UE particles in the cone. The perpendicular cone particle source (PC) consists of all particles inside $R=0.4$ cones at $\Delta \varphi =90^{\circ}$ and $\Delta \eta =0$ from the selected anti-$k_{\rm{T}}$ jet cones. PC particles are expected to be uncorrelated with the hard scattering, so they are used to perform UE subtraction, following Ref.~\cite{CascadespPbpp}. The inclusive particle production (without any jet requirements) is also measured to compare UE particle production to inclusive particle production.

PID is performed via fits to the Time Of Flight (TOF) $n_\sigma$ particle hypothesis distributions. These fits are performed separately for each particle source and each $\pi$, K, p species hypothesis. The raw particle yield as a function of $\textit{p}_{\mathrm{T}}$ is then obtained by integrating these fits for each species. Standard $\textit{p}_{\mathrm{T}}$-dependent, bin-by-bin PID corrections (tracking efficiency, TOF matching efficiency, and primary fraction) are performed using Monte Carlo simulations.

The perpendicular cone method is used to estimate and remove the UE constituents from the JC constituents. To do this, the yields are normalized by area, i.e.

\begin{equation}
\frac{\rm{d}\rho}{\rm{d}\textit{p}_{\mathrm{T}}^{\rm{track}}} = \frac{1}{N_{\rm{trig~evt}}} \frac{1}{A_{\rm{acc}}} \frac{\rm{d}\it{N}}{\rm{d}\textit{p}_{\mathrm{T}}^{\rm{track}}},
\label{eq:density}
\end{equation}
where $N_{\rm{trig~evt}}~A_{\rm{acc}}$ is the total acceptance area for triggered events for each particle source, and $\frac{\rm{d}\it{N}}{\rm{d}\textit{p}_{\mathrm{T}}^{\rm{track}}}$ is the particle yield. For inclusive particles ($\frac{\rm{d}\rho_{\rm{inc}}}{\rm{d}\textit{p}_{\mathrm{T}}^{\rm{track}}}$), $N_{\rm{trig~evt}}~A_{\rm{acc}}$ is the entire ALICE tracking area (1.8 in $\eta$ $\times$ 2$\pi$ in $\varphi$) times the number of selected events. For JC particles ($\frac{\rm{d}\rho_{\rm{JC}}}{\rm{d}\textit{p}_{\mathrm{T}}^{\rm{track}}}$), $N_{\rm{trig~evt}}~A_{\rm{acc}}$ is the total area of all jet cones from all jet events. For PC particles ($\frac{\rm{d}\rho_{\rm{PC}}}{\rm{d}\textit{p}_{\mathrm{T}}^{\rm{track}}}$), $N_{\rm{trig~evt}}~A_{\rm{acc}}$ is the number of perpendicular cones times $\pi$$R^2$ where $R=0.4$. The PC yield is used both to compare the jet and UE yields and to subtract the UE from the JC.

The performance of directly subtracting $\frac{\rm{d}\rho_{\rm{PC}}}{\rm{d}\textit{p}_{\mathrm{T}}^{\rm{track}}}$ from $\frac{\rm{d}\rho_{\rm{JC}}}{\rm{d}\textit{p}_{\mathrm{T}}^{\rm{track}}}$ is checked in a toy model of PYTHIA8 jets embedded into a thermal background \cite{PYTHIA8}. In this model, the number of PC particles slightly underestimates the number of thermal particles in the jets. At low $\textit{p}_{\mathrm{T}}$, the thermal particles in the PYTHIA+Thermal jets dominate the PYTHIA particles in these jets. So, even slightly underestimating the number of thermal particles led to UE subtracted jet spectra that were far off from the PYTHIA truth. The difference between the PC spectra and the jet thermal background is due to a jet-selection bias.

This jet-selection bias occurs because the jets are required to have a minimum reconstructed jet $\textit{p}_{\mathrm{T}}$ after the pedestal subtraction (60 GeV/$c$ here). Because jet spectra are steeply falling, jets reconstructed in the presence of the thermal background tend to originate from a PYTHIA jet with a jet $\textit{p}_{\mathrm{T}}$ just below 60 GeV/$c$ that gets pushed to over 60 GeV/$c$ after the pedestal subtraction due to sitting on an upward fluctuation in the background. More jets are produced with a jet $\textit{p}_{\mathrm{T}}$ below 60 GeV/$c$ than above it, so the jets sitting on upward fluctuations dominate over those on downward fluctuations. These effects cause the yield of particles in the PC to be lower than the yield of thermal particles in the PYTHIA+Thermal jets. 

This effect can be corrected by applying a scaling factor to the yields measured in the perpendicular cones, obtained from PYTHIA events embedded into ALICE 0--10\% minimum-bias Pb--Pb data. This scaling factor is given by 

\begin{equation}
c_{\rm{UE~bias}}(\textit{p}_{\mathrm{T}}^{\rm{track}}) = \frac{\rm{d}\rho_{\rm{UE~in ~selected~jets}}}{\rm{d}\textit{p}_{\mathrm{T}}^{\rm{track}}} / \frac{\rm{d}\rho_{\rm{PC}}}{\rm{d}\textit{p}_{\mathrm{T}}^{\rm{track}}}.
\end{equation}
The numerator contains the \mbox{Pb--Pb} background particles found in embedded jets, i.e. the UE as seen by the jet. The denominator contains the particles in the cones perpendicular to these embedded jets. Separate scaling factors are obtained for each particle species to account for any possible species dependence. Finally, the jet signal particle densities are given by

\begin{equation}
\frac{\rm{d}\rho_{\rm{JE}}}{\rm{d}\textit{p}_{\mathrm{T}}^{\rm{track}}} = \frac{\rm{d}\rho_{\rm{JC}}}{\rm{d}\textit{p}_{\mathrm{T}}^{\rm{track}}} - \frac{\rm{d}\rho_{\rm{PC}}}{\rm{d}\textit{p}_{\mathrm{T}}^{\rm{track}}}*c_{\rm{UE~bias}}.
\end{equation}
These results include a conservative systematic of applying this scaling factor as described compared to trivially setting $c_{\rm{UE~bias}}=1$ to account for possible contamination from jets in the ALICE minimum-bias events used for embedding. 

\section{Results}
\label{results}

\begin{figure}[!h]
\centering
\includegraphics[width=10.2cm,clip]{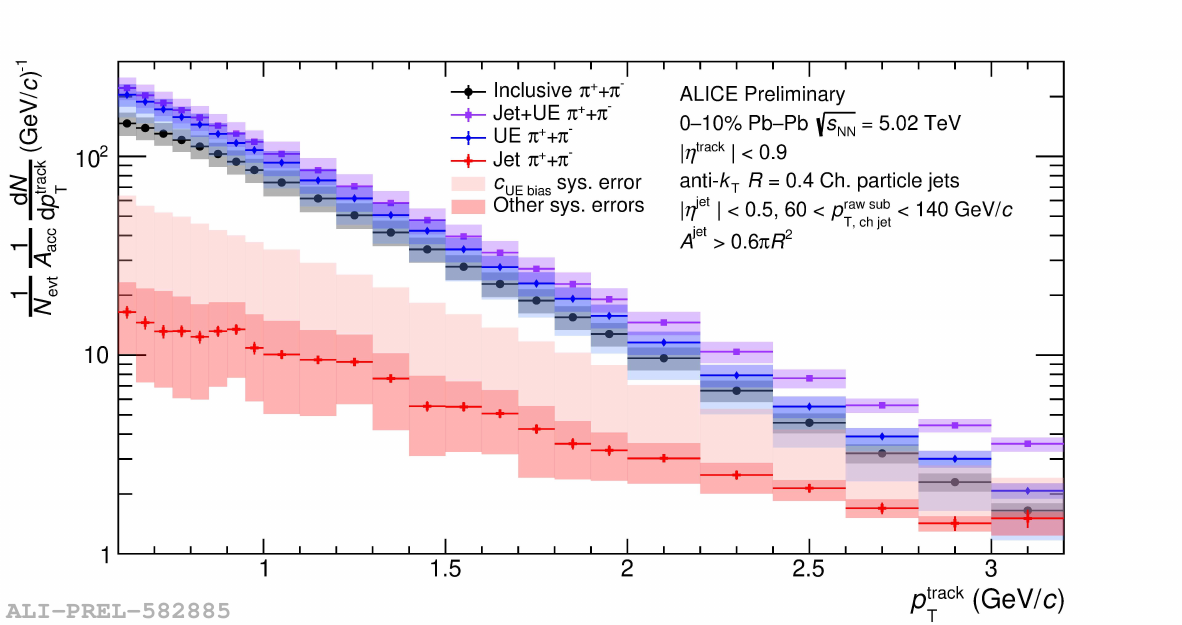}
\caption{$\pi$ spectra for inclusive, jet+UE, UE, and jet signal in 0$-$10\% \mbox{Pb--Pb} collisions.}
\label{fig:pi yields}
\end{figure} 

\begin{figure}[!h]
\centering
\includegraphics[width=10.2cm,clip]{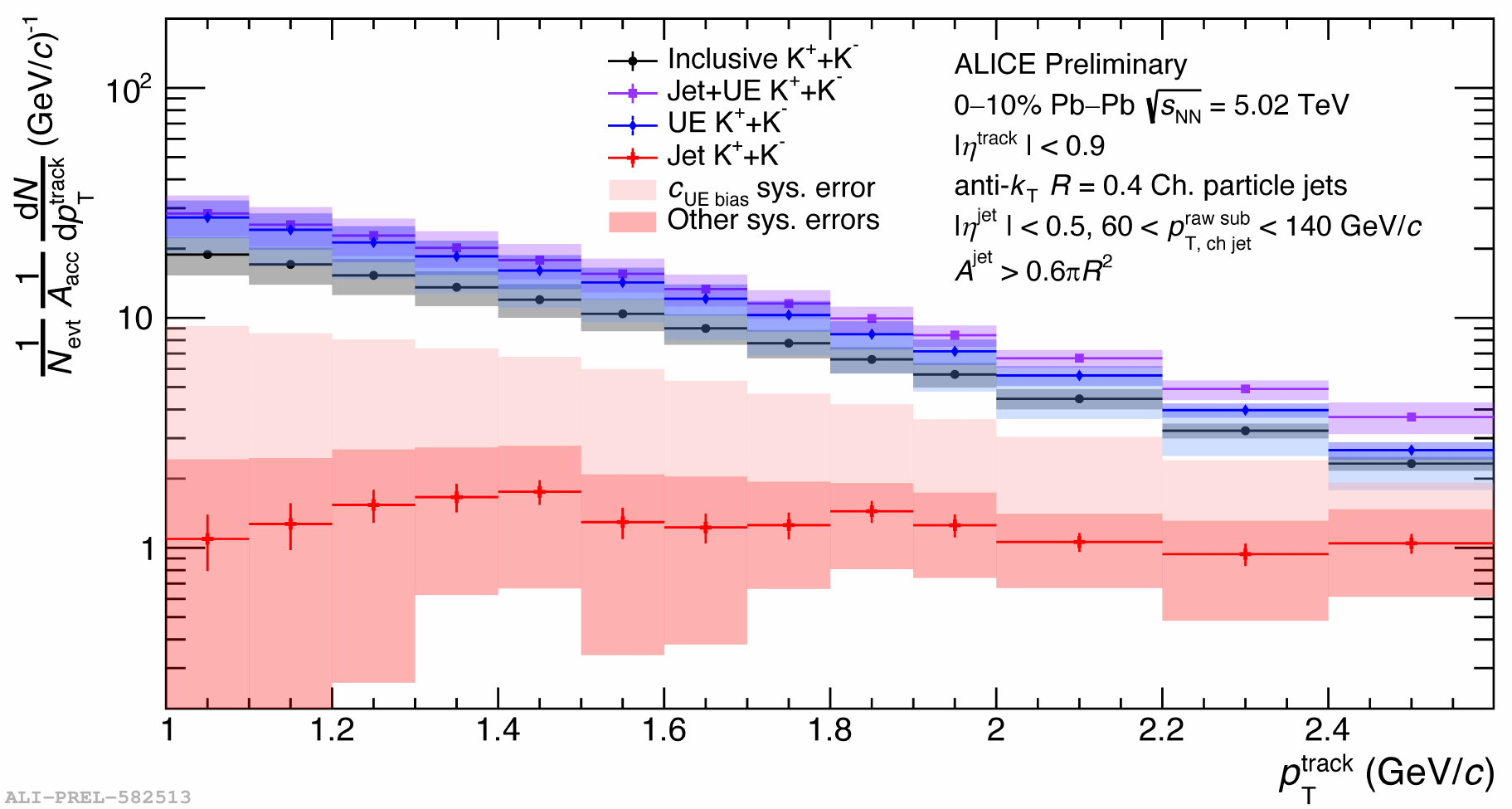}
\caption{K spectra for inclusive, jet+UE, UE, and jet signal in 0$-$10\% \mbox{Pb--Pb} collisions.}
\label{fig:K yields}
\end{figure} 

\begin{figure}[h]
\centering
\includegraphics[width=10.2cm,clip]{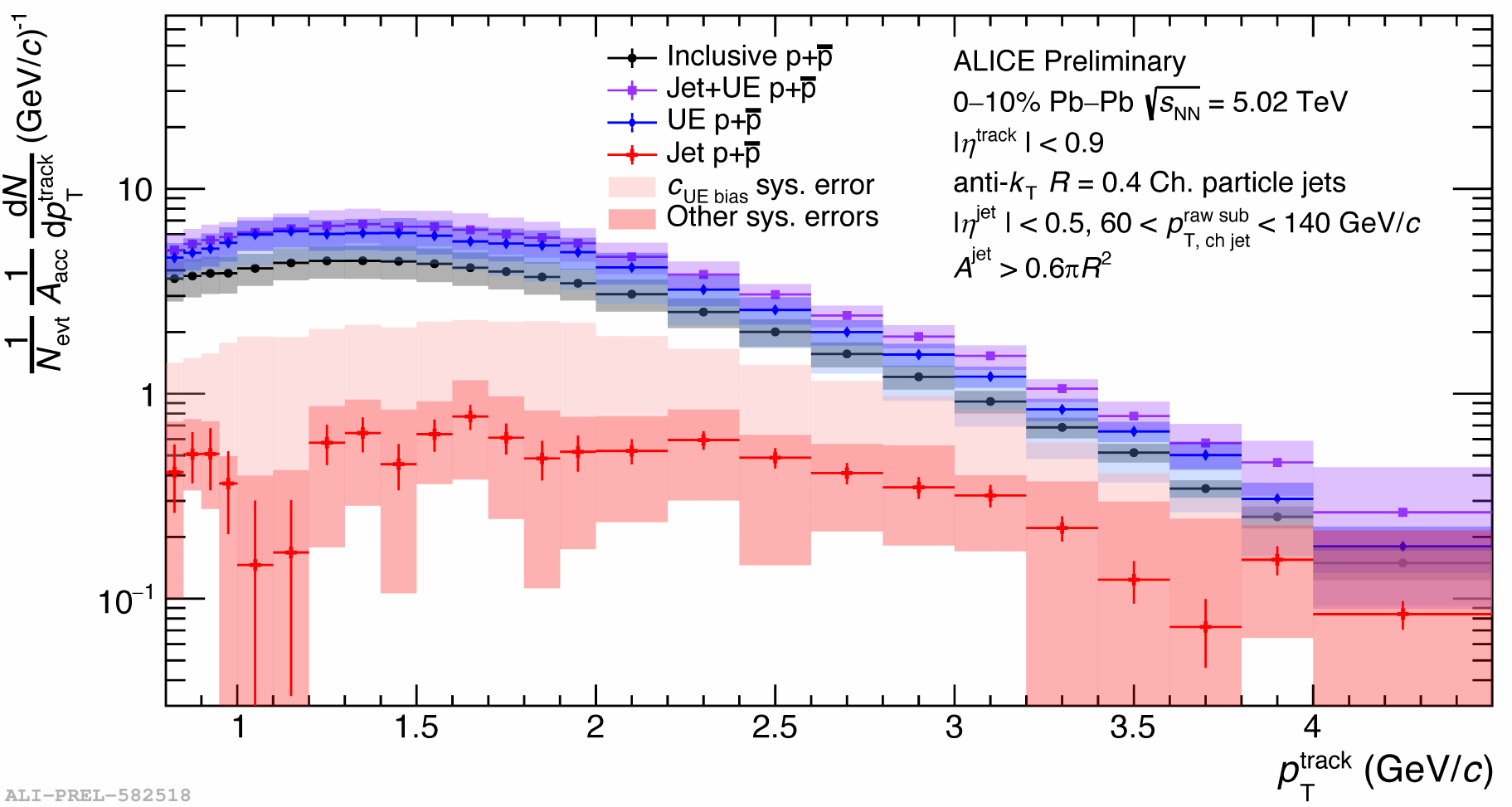}
\caption{p spectra for inclusive, jet+UE, UE, and jet signal in 0$-$10\% \mbox{Pb--Pb} collisions.}
\label{fig:p yields}
\end{figure}

The \mbox{Pb--Pb} corrected spectra for each particle source are shown in Figures \ref{fig:pi yields}\textendash\ref{fig:p yields}. The inclusive and jet+UE spectra are given directly by $\frac{\rm{d}\rho_{\rm{inc}}}{\rm{d}\textit{p}_{\mathrm{T}}^{\rm{track}}}$ and $\frac{\rm{d}\rho_{\rm{JC}}}{\rm{d}\textit{p}_{\mathrm{T}}^{\rm{track}}}$, respectively. The UE spectra is given by 
\begin{equation}
\frac{\rm{d}\rho_{\rm{UE}}}{\rm{d}\textit{p}_{\mathrm{T}}^{\rm{track}}} = \frac{\rm{d}\rho_{\rm{PC}}}{\rm{d}\textit{p}_{\mathrm{T}}^{\rm{track}}} c_{\rm{UE~bias}},
\end{equation}
and the JE spectra is given by

\begin{equation}
\frac{\rm{d}\rho_{\rm{JE}}}{\rm{d}\textit{p}_{\mathrm{T}}^{\rm{track}}} = \frac{\rm{d}\rho_{\rm{Jet+UE}}}{\rm{d}\textit{p}_{\mathrm{T}}^{\rm{track}}} - \frac{\rm{d}\rho_{\rm{UE}}}{\rm{d}\textit{p}_{\mathrm{T}}^{\rm{track}}}.
\end{equation} 
For all species, the UE and Pb--Pb inclusive yields are consistent within uncertainties in the reported $\textit{p}_{\mathrm{T}}$ range. The Jet+UE is dominated by the UE for all species in the $\textit{p}_{\mathrm{T}}$ range considered, though the jet portion of the jet+UE yields gets fractionally larger as $\textit{p}_{\mathrm{T}}$ increases.

  \begin{figure}[!h]
  \centering
  \includegraphics[width=10.0cm,clip]{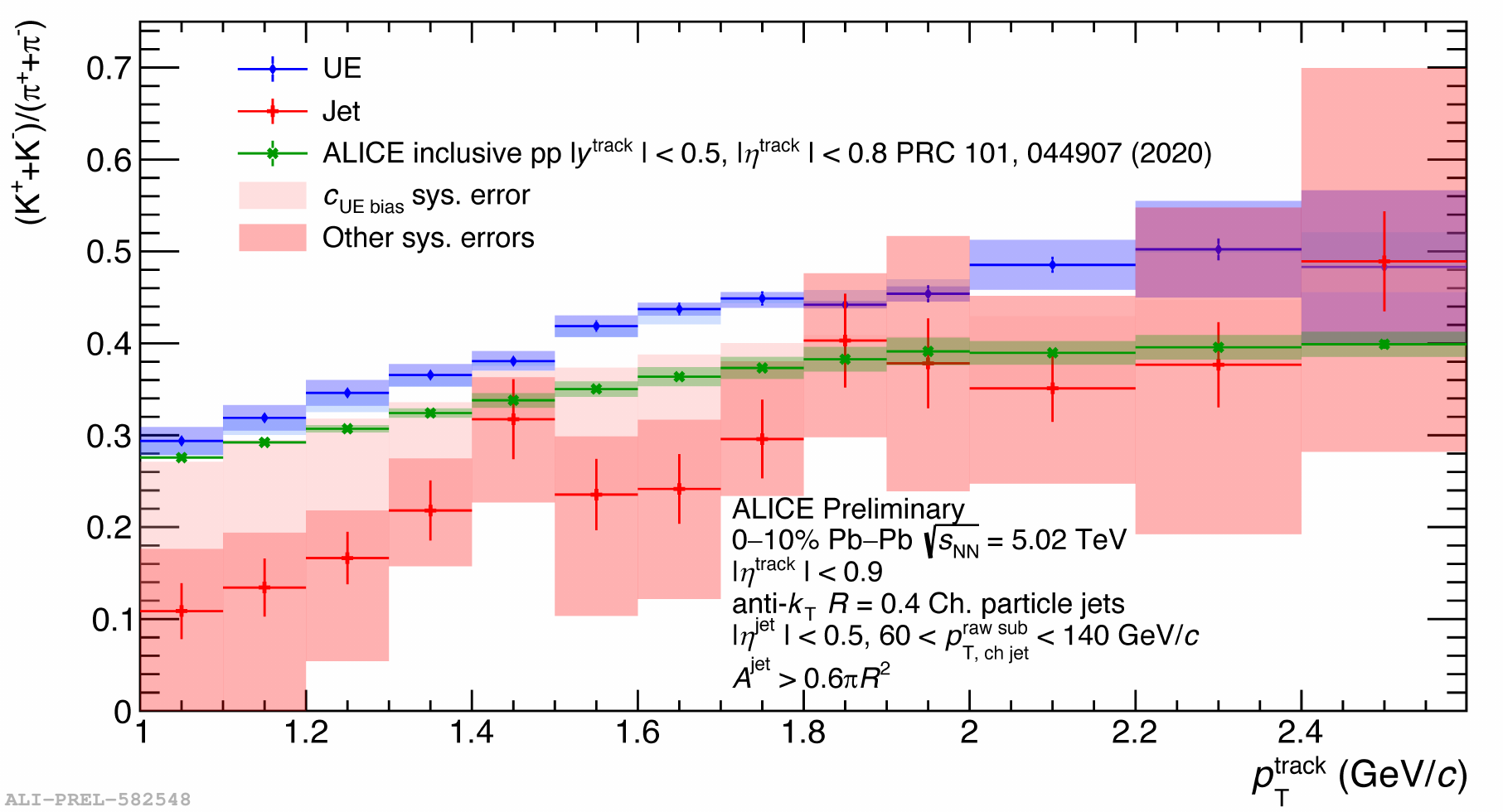}
  \includegraphics[width=10.0cm,clip]{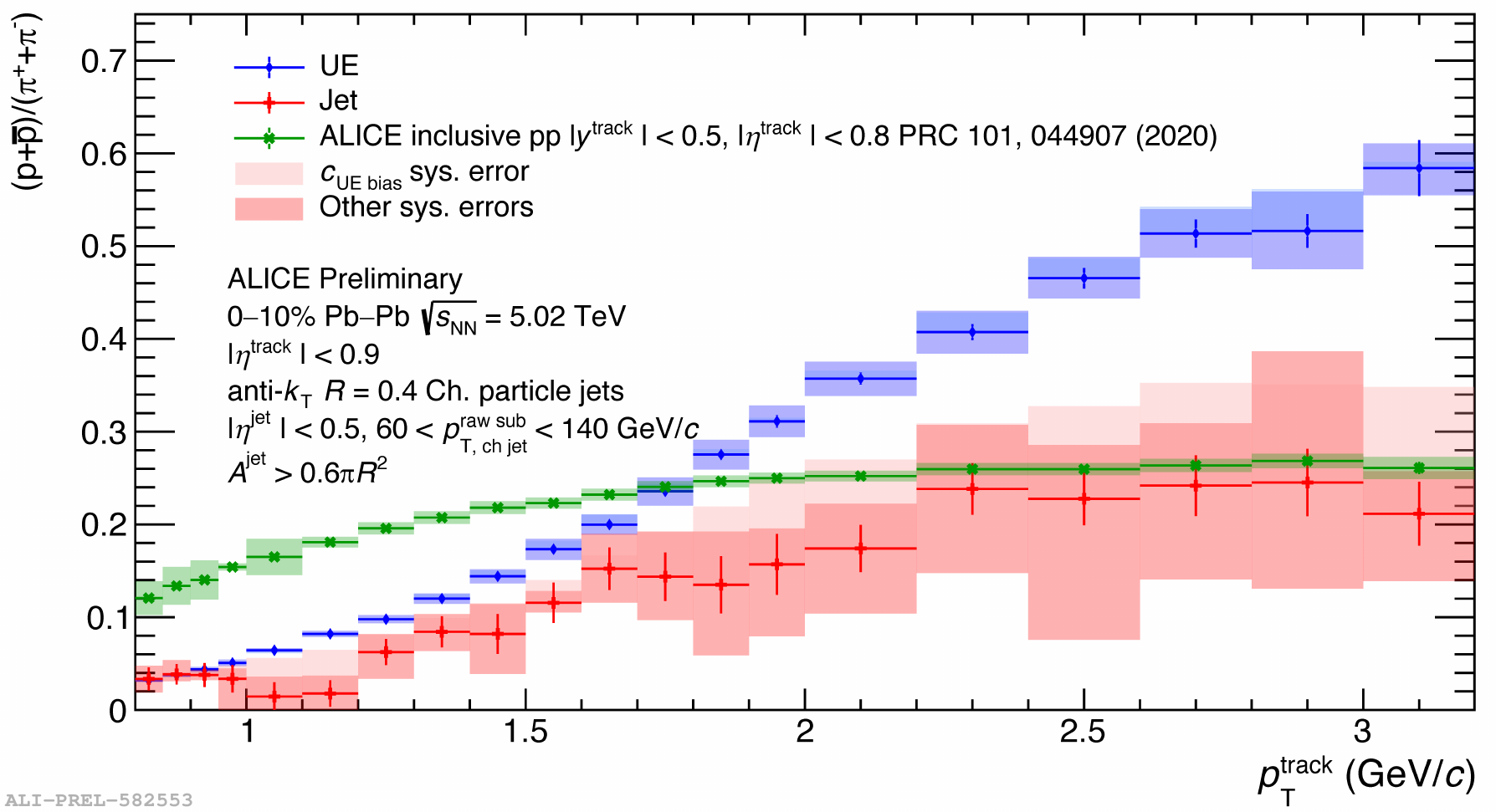}
\caption{K/$\pi$ (top) and p/$\pi$ (bottom) particle ratios in jets and the UE as a function of particle $\textit{p}_{\mathrm{T}}$ for $R$=0.4 jets with $60 < p^{\rm{raw~sub}}_{\rm{T,~ch~jet}} < 140 $ GeV/$c$. }
\label{Sierra_piKp_ratios}
\end{figure}

  The species ratios are shown in Figure \ref{Sierra_piKp_ratios}. The p/$\pi$ ratio is lower in jets than in the UE, particularly at intermediate $\textit{p}_{\mathrm{T}}^{\rm{track}}$ (2 \textendash~3.2 GeV/$c$). This shows that there is less baryon production relative to meson production in jets compared to the UE. The jet p/$\pi$ ratio is also lower than the p/$\pi$ ratio in inclusive pp at low $\textit{p}_{\mathrm{T}}^{\rm{track}}$. The jet K/$\pi$ ratio is systematically lower than both the K/$\pi$ ratio in \mbox{Pb--Pb} UE and pp inclusive, hinting at less strangeness production in \mbox{Pb--Pb} jets, but the systematic uncertainties are too large to make definitive conclusions. Though these results are compared to inclusive pp measurements, the inclusive pp particle ratios are expected to differ from the pp jet particle ratios \cite{CascadespPbpp}. Thus, future measurements of these ratios in pp jets are necessary to probe possible jet hadrochemistry modifications.

\section{Conclusions}
\label{conclusions}

These proceedings present the first measurement of $\pi$, K, p production in jets and UE in \mbox{Pb--Pb} collisions. The p/$\pi$ ratio is lower in jets than in the UE, indicating that baryon production relative to meson production in \mbox{Pb--Pb} jets is less than the \mbox{Pb--Pb} UE. The measured K/$\pi$ ratios in \mbox{Pb--Pb} UE and \mbox{Pb--Pb} jets hint at less strangeness production in \mbox{Pb--Pb} jets than \mbox{Pb--Pb} UE. Performing pp jet K/$\pi$ and p/$\pi$ measurements is essential to probe possible jet hadrochemistry modification due to modified fragmentation or medium response. Planned extensions of this analysis include performing the measurement in pp at $\sqrt{s}=5.02$ TeV, reducing the Pb--Pb jet systematic uncertainties, unfolding these results to enable robust comparisons to theory and investigations of possible jet $\textit{p}_{\mathrm{T}}$ dependence, extending the PID particle $\textit{p}_{\mathrm{T}}$ range of these results by including the TPC PID capabilities, and investigating the centrality dependence of the particle ratios. Finally, a multi-differential analysis of the radial dependence of the jet K/$\pi$ and p/$\pi$ ratios is planned, as the wake response is expected to further increase the jet hadrochemistry modification at larger angles \cite{Luo:2023}.

%
\bibliography{refs.bib}

\end{document}